\documentclass {aipproc}
\layoutstyle{8x11single}

\SetInternalRegister\hbadness{8000} % pseudo latin isn't breaking very well :-)

\newcommand\doingARLO[2][]{%
  \ifx\mmref\undefined #1\else #2\fi
}

\begin{document}

\title 
      {The Isoscalar Giant Dipole Resonance in $^{208}$Pb and the Nuclear Incompressibility}
%\classification{43.35.Ei, 78.60.Mq}
%\keywords{Document processing, Class file writing, \LaTeXe{}}

\author{M.~Hedden}{
  address={Physics Department, University of Notre Dame, Notre Dame, IN 46556},
}

\author{U.~Garg}{
  address={Physics Department, University of Notre Dame, Notre Dame, IN 46556},
}

\author{B.~Kharraja}{
  address={Physics Department, University of Notre Dame, Notre Dame, IN 46556},
}

\author{S.~Zhu}{
  address={Physics Department, University of Notre Dame, Notre Dame, IN 46556},
}

%\iftrue
\author{M.~Uchida}{
  address={Department of Physics, Kyoto University, Kyoto 606-8502, Japan},
}

\author{H.~Sakaguchi}{
  address={Department of Physics, Kyoto University, Kyoto 606-8502, Japan},
}

\author{T.~Murakami}{
  address={Department of Physics, Kyoto University, Kyoto 606-8502, Japan},
}

\author{M.~Yosoi}{
  address={Department of Physics, Kyoto University, Kyoto 606-8502, Japan},
}

\author{H.~Takeda}{
  address={Department of Physics, Kyoto University, Kyoto 606-8502, Japan},
}

\author{M.~Itoh}{
  address={Department of Physics, Kyoto University, Kyoto 606-8502, Japan},
}

\author{T.~Kawabata}{
  address={Department of Physics, Kyoto University, Kyoto 606-8502, Japan},
}

\author{T.~Taki}{
  address={Department of Physics, Kyoto University, Kyoto 606-8502, Japan},
}

\author{T.~Ishikawa}{
  address={Department of Physics, Kyoto University, Kyoto 606-8502, Japan},
}

\author{N.~Tsukuhara}{
  address={Department of Physics, Kyoto University, Kyoto 606-8502, Japan},
}

\author{Y.~Yasuda}{
  address={Department of Physics, Kyoto University, Kyoto 606-8502, Japan},
}

\author{M.~Fujiwara}{
  address={Research Center for Nuclear Physics, Osaka University, Suita 567, Japan},
}

\author{H.~Fujimura}{
  address={Research Center for Nuclear Physics, Osaka University, Suita 567, Japan},
}

\author{H.P.~Yoshida}{
  address={Research Center for Nuclear Physics, Osaka University, Suita 567, Japan},
}

\author{E.~Obayashi}{
  address={Research Center for Nuclear Physics, Osaka University, Suita 567, Japan},
}

\author{K.~Hara}{
  address={Research Center for Nuclear Physics, Osaka University, Suita 567, Japan},
}

\author{H.~Akimune}{
  address={Department of Physics, Konan University, Kobe 658-8501, Japan},
}
\author{M.N.~Harakeh}{
  address={KVI, 9747 AA Groningen, The Netherlands},
}
\author{M.~Volkerts}{
  address={KVI, 9747 AA Groningen, The Netherlands},
}
%\fi
%\begin{center}
%{Invited talk at INPC2001, Berkeley, CA, July 30-Aug. 3, 2001 (to be published
%by AIP)}
%\end{center}
% \copyrightholder{Acoustical Scociety of America}
%\copyrightyear  {2001}

\begin{abstract}
The {\em isoscalar} giant dipole resonnace (ISGDR) has been investigated in
$^{208}$Pb using inelastic scattering of 400 MeV $\alpha$ particles at 
extremely forward angles, including 0$^{\circ}$. Using the superior
capabilities of the Grand Raiden spectrometer, it has been possible to obtain
inelastic spectra devoid of any ``instrumental'' background. The ISGDR
strength distribution has been extracted from a multipole-decomposition of the
observed spectra. The implications of these results on the experimental value 
of nuclear incompressibility are discussed.
\end{abstract}

%\date{\today}

\maketitle

%\section{Introduction}
Nuclear Incompressibility is a crucial component of the nuclear equation of 
state and, as such, has very important bearing on diverse nuclear and 
astrophysical phenomena:
for example, strength of supernova collapse, the emission of neutrinos in
supernova explosions, and collective flow in medium- and high-energy heavy-ion
collisions. The only direct experimental measurement of nuclear 
incompressibility is possible via the compressional-mode giant resonances in
nuclei. Of the two important comprssional modes, the isoscalar giant monopole
resonance (GMR) is the better known and has been studied
now for more than 20 years. The other mode, the {\em isoscalar} giant dipole
resonance
(ISGDR), is an exotic oscillation---to first order, isoscalar dipole mode 
represents simply the motion of the center-of-mass which, in itself, cannot 
lead to any
nuclear excitation---which has remained somewhat elusive, even though initial
(although inconclusive) evidence for the mode was reported as long ago as the 
early 80's. 

The ISGDR can be thought of a hydrodynamic density oscillation in which the 
volume of the nucleus remains constant and a compressional wave---akin to a
sound wave---traverses back and forth through the nucleus. The mode has 
generally been referred to as the ``squeezing mode'' in analogy with the
mnemonic ``breathing mode'' for the GMR. A general description of this mode,
along with the relevant transition densities and sum-rules, has been provided 
in Refs. \cite{hara1,hara2}.
The energy of this resonance is
related to the nuclear incompressibilty via the scaling relation:

\begin{equation}
E_{ISGDR} = \hbar \sqrt{\frac{7}{3} \frac{K_{A}+\frac{27}{25}\epsilon
_{F}}{m< r^{2} >}},
\end{equation}

\noindent
where K$_{A}$ is the incompressibility of the nucleus, $m$ is the nucleon mass,
and $\epsilon _{F}$ is the Fermi energy \cite{strin}.

As mentioned above, indications of the ISGDR were reported as early as the
beginning of the 1980's. However, the first conclusive evidence for this mode,
based on the differences in angular distribution of the ISGDR from that of the 
nearby high-energy octupole resonance (HEOR), was provided by Davis
{\em et al.} \cite{Davis}, who
demonstrated that in inelastic scattering of 200 MeV $\alpha$'s at angles near 
0$^{\circ}$, the giant resonance ``bump'' at
3 $\hbar \omega$ excitation energy could be separated into two components, with 
the higher-energy component corresponding to the ISGDR. Further evidence for the ISGDR
has since come from 240 MeV $\alpha$ inelastic scattering measurements, using
the multipole-decomposition technique \cite{hen,dhy}. For reviews of the status
of the ISGDR work, the reader is referred to Refs. \cite{Garg98,Garg99,Garg99b}.

We have undertaken a detailed investigation of this resonance using inelastic
scattering of 400 Mev $\alpha$'s at forward angles, including 
0$^{\circ}$. The experiments are being carried out at the Research Center for
Nuclear Physics (RCNP), Osaka University. The ring-cyclotron 
facility at RCNP provided high-quality $\alpha$-particle beams of 400 MeV energy
which were momentum analyzed in a recently-installed beam-analysis system and 
were incident on
thin (typically 2--10 mg cm$^{-2}$), self-supporting, metallic targets. Data
were obtained on $^{58}$Ni, $^{90}$Zr, $^{116}$Sn, $^{144,148,150,152,154}$Sm, 
and $^{208}$Pb over several angles between 0$^{\circ}$ and 13$^{\circ}$. In 
addition, spectra were measured for $^{12}$C targets for 
energy-caliberation purposes. In this report, we will concentrate on the
results for $^{208}$Pb; preliminary results on the 
Sm isotopes have been presented previously \cite{itoh}.
  
Inelastically scattered particles were analyzed by a versatile, high-resolution,
magnetic spectrometer, Grand Raiden \cite{fuji}. The focal-plane
detector system \cite{tam,noro} consists of two
multi-wire drift chambers (MWDC) and two plastic
scintillation counters. The MWDC's provided the position and angle information 
which was used, by ray-tracing technique, to subdivide the full
angular opening of the spectrometer ($\sim$2$^{\circ}$) into several angular
bins and to construct energy spectra for the corresponding scattering angles.
Particle identification was provided by 
the energy-loss signals from the plastic scintillators. For the 0$^{\circ}$
measurements, the primary beam, after passing through the spectrometer, was
guided through a hole in the high-momentum side of the detector system, and
was stopped in a Faraday cup placed several meters downstream from the
focal plane. For the extremely forward angles (2$^{\circ}$--5$^{\circ}$), the 
beam was stopped just behind the first quadrupole magnet of Grand Raiden (about
160 cm downstream from the target); 
for all other angles, a Faraday cup was placed inside the target chamber.

Small-angle measurements, as is well-known, require extra care in 
beam-preparation and transport through the system. Several innovative measures
were
adopted to ensure a halo-free beam and, ultimately, the ``blank-target'' rate 
in the focal-plane detector system was typically only $\sim$10/second, compared 
with count rates of several thousand per second with the targets in place. 
The typical energy resolution was $\sim$250 keV, more than sufficient to
investigate the giant resonances which are typically several-MeV wide. 

\vspace{-3.5cm}
\begin{figure}[ht]
    \resizebox{23pc}{!}{\includegraphics{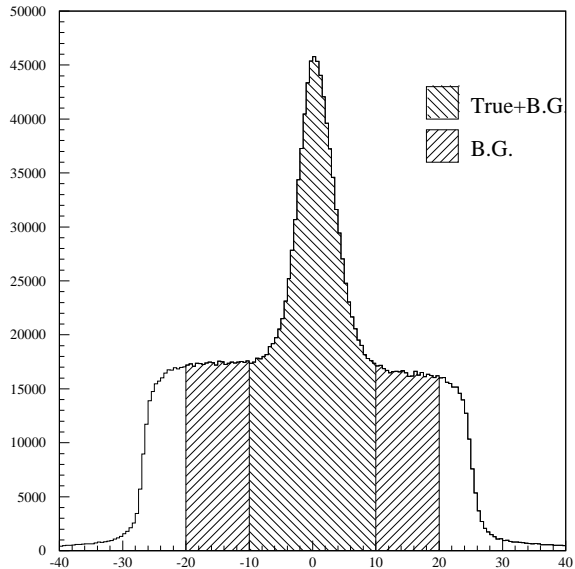}}
\vspace{-0.3cm}
\caption{Vertical ($y$) position spectrum in the ``0$^{\circ}$'' inelastic
scattering measurement
on $^{208}$Pb. The peak represents ``true'' events; the flat part emanates from
non-physical (``background'') events.}
\end{figure}

A very important aspect of these measurements has been the  elimination of 
all ``instrumental
background'' from the final spectra. This ``background'' results primarily from 
multiple scattering of the beam on the
target, as well as from rescattering by the yoke and the walls of the
spectrometer. For the purpose, we utilized the vertical-position spectrum at 
the 
double-focused position of the spectrometer. As shown in Fig. 1, the ``true''
events (those coming from
inelastically scattered $\alpha$ particles) focus well at the
focal plane due to  the ion-optics of the spectrometer (the peak in Fig. 1);
the ``background'' events, on the other hand, have a poor focus and the 
distributions of the vertical positions of such events is ``flat''.
The ``background'' events were, then, subtracted from the final spectra by
gating on the peak in Fig. 1 and subtracting the events corresponding to the
underlying background which can be estimated by setting gates on both sides of 
the peak.

Fig. 2 shows the results of these cuts. The top left part of the figure shows
the spectrum for $^{208}$Pb at ``0$^{\circ}$'' (the actual angle is
0.77$^{\circ}$) resulting from a gate on the peak in Fig. 1. The top right part
shows the spectrum with gate on the background. In the bottom part 
of the figure is shown the spectrum with the background subtracted. As noted
above, this ability to eliminate all instrumental background from singles 
inelastic scattering spectra has been achieved for the first time in such
studies and has led to observation of GR strength in the excitation-energy
region up to $\sim$30 MeV; in previous measurements, all this strength at the
higher excitation energies has,
generally, been part of the overall empirical background subtraction. The
``flat'' part at the upper end of the excitation-energy spectrum, we believe, 
is the true physical continuum.

\vspace{-1cm}
\begin{figure}[ht]
\includegraphics[height=15cm]{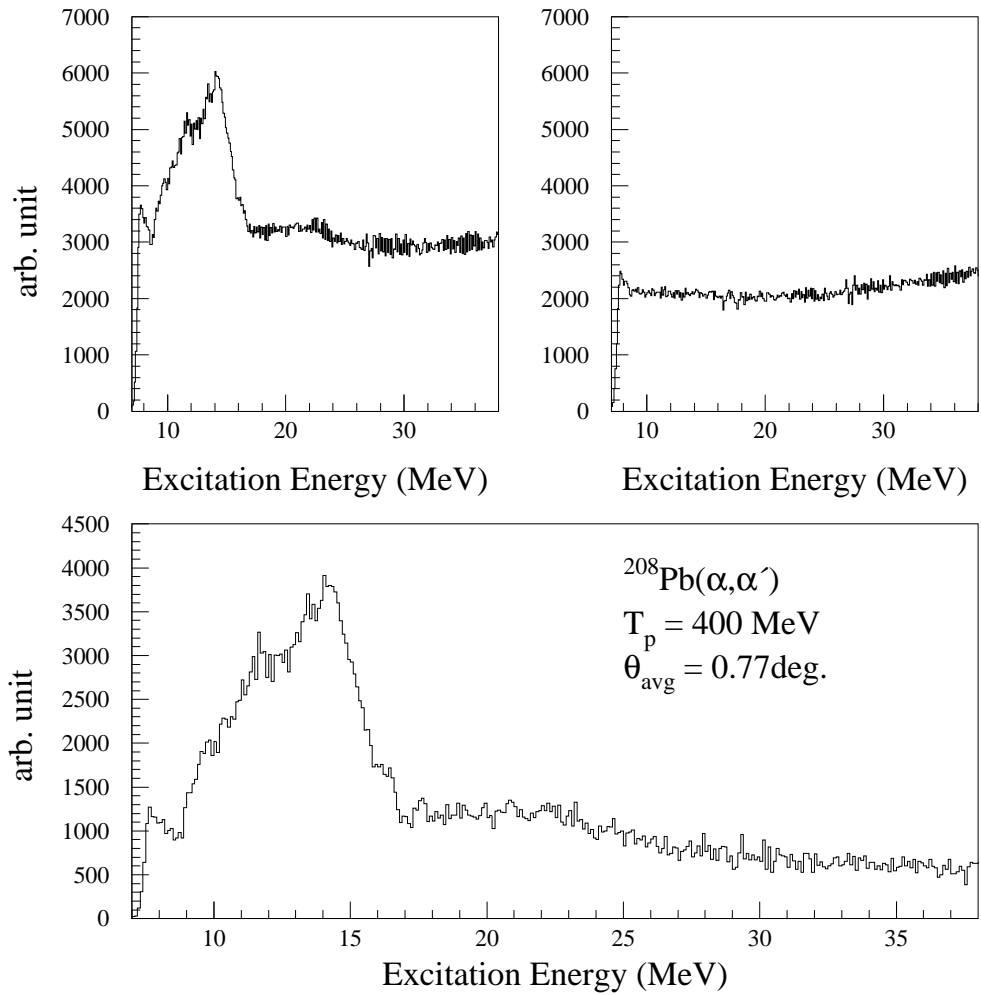}
\caption{(Top Left): Position spectrum gated on the ``peak'' in Fig. 1. (Top
Right): Position spectrum gated on  the corresponding ``background''. (Bottom):
``Background-free'' position spectrum obtained by subtracting the Top Right
spectrum from the Top Left spectrum.}
\end{figure}

The strength distributions for the various modes has been extracted from these
spectra using a multipole-decomposition analysis, similar to that used
previously by Bonin {\em et al.} \cite{bonin} and, more recently, also by 
Youngblood~{\em et al.} \cite{dhy}. The contribution of the isovector giant 
dipole resonance (IVGDR), which is excited because of Coulomb interaction, is 
subtracted from the spectra, using the precisely-known IVGDR strength
distributions from previous photoabsorption work \cite{berman}. In this 
analysis, the ``continuum'' has not been subtracted out; instead, the entire
spectrum has been fitted with a combination of multipoles (up to $l$=5;
the available angular-range for these measurements is not sufficient to clearly
distinguish the higher-order multipoles from those already employed).

\vspace{0.5cm}
\begin{figure}[ht]
\includegraphics[height=9.0cm]{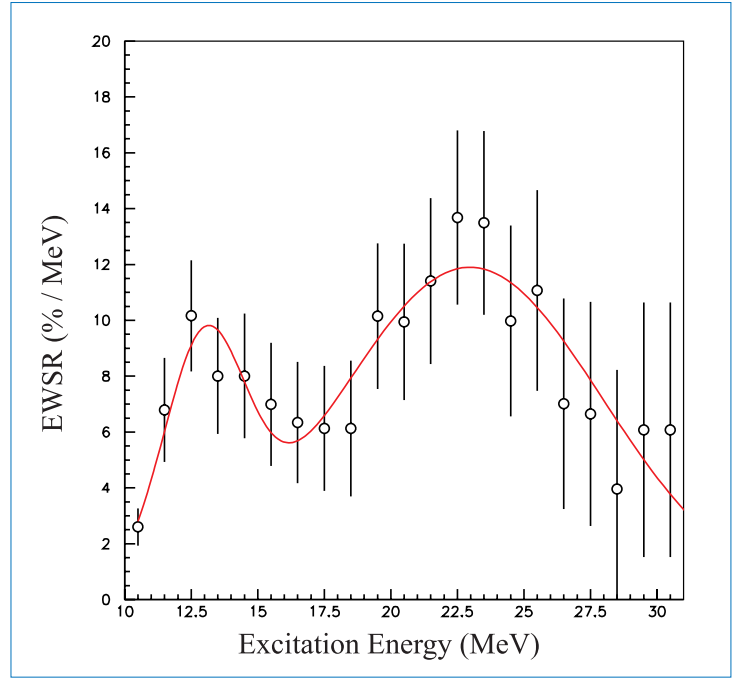}
\caption{Preliminary ISGDR strength distribution (open circles) in $^{208}$Pb, 
as obtained in 
the present work. The solid line represents a 2-Gaussian fit to the 
distribution.}
\end{figure}

The resulting ISGDR strength distribution is shown in Fig. 3. While these 
results must be considered preliminary at this stage, pending complete data
analysis, the observed ISGDR distribution clearly has two components, centered
at about 13 MeV and 23 MeV,
respectively. Results of a two-Gaussian fit to the strength distribution are
shown superimposed in the figure and are presented in Table 1.\footnote{The 
errors quoted for the present work are primarily statistical. There are 
additional uncertainties associated with extraction of ISGDR strength from DWBA 
calculations.} Also shown in Table 1 are the results from other recent 
measurements on ISGDR as well as the theoretical results from various recent 
calculations.

%\vspace{-1cm}
This ``bi-modal'' ISGDR strength distribution is most inetresting and, as
discussed below, has crucial bearing on the determination of nuclear
incompressibility using the excitation energy of the ISGDR. The 
lower-energy component, at about the energy of the GMR and IVGDR, has now been
observed in several nuclei 
and also appears in several of the more recent calculations of
ISGDR strength (see, for example, Refs. \cite{colo,dario,jorge}).\footnote{Incidentally, this is not the
expected 1$\hbar \omega$ component of the ISGDR, disussed in detail previously
by Poelhekken {\em et al.} \cite{poel}. The 1$\hbar \omega$ strength would be
expected near or below the lower-excitation-energy bound of our experiment.} 
All the cited 
calculations clearly establish that only the higher-energy component depends on
the value of nuclear incompressibility employed in the calculations; the 
position of the lower-energy component is completely independent of the
K$_{\infty}$'s, pointing to its ``non-bulk'' origin. Further, Vretenar
{\em et al.}
\cite{dario} have identified the dynamics of this mode as resulting from
surface effects. We note, however, that the observed centroid for this
low-energy component is higher than the theoretical predictions by several MeV.

%\vspace{1cm}
\begin{table}[ht]
\begin{tabular}{lllllllll}
  \\
\hline
  & \tablehead{1}{l}{b}{     }
  & \tablehead{1}{l}{b}{     }
  & \tablehead{1}{l}{b}{     }
  & \tablehead{1}{l}{b}{High-energy\\component}
  & \tablehead{1}{l}{b}{     }
  & \tablehead{1}{l}{b}{     }
  & \tablehead{1}{l}{b}{     }
  & \tablehead{1}{l}{b}{Low-energy\\component}   \\
\hline
This work & & & & 23.0 $\pm$ 0.5 MeV & & & & 13.0 $\pm$ 0.5 MeV\\
%\tablenote{The quoted errors are upper-bound estimates}
Davis {\em et al.}\tablenote{Ref. \cite{Davis}} & & & & 22.4 $\pm$ 0.5 & & & & \\
Youngblood {\em et al.}\tablenote{Ref. \cite{dhy}} & & & & 19.9 $\pm$ 0.8 & & & & 12.2 $\pm$ 0.6\\
Colo {\em et al.}\tablenote{RPA with SGII, K$_{\infty}$ = 215 MeV \cite{colo}} && & &  23.9 & & & & 10.9\\
Vretenar {\em et al.}\tablenote{RRPA with NL3, K$_{\infty}$ = 271.8 MeV \cite{dario}} & & & & 26.01 & & & & 10.4\\
Piekarewicz\tablenote{RRPA with NL-C, K$_{\infty}$ = 224 MeV \cite{jorge,jorge2}} & & & & 24.4 & & & & $\sim$8\\
\hline
\end{tabular}
\caption{ISGDR excitation-energies obtained in the present work. Also shown are
results from other recent measurements and predictions from recent theoretical
work.}
\label{tab:a}
%\vspace{-1.5cm}
\end{table}

%\vspace{-1cm}

The extraction of a value for the nuclear incompressibility using 
simultaneously the excitation energies of the two compressional modes, the GMR
and ISGDR, has been problematic so far \cite{Garg99b}. Until recently, it
appeared that all the calculations, using nuclear incompressibility that
reproduced the GMR energies well, substantially overpredicted the excitation 
energy of the ISGDR.
With the results reported herein, this ``problem''
has been alleviated to a great extent. First, with the ``background-free'' 
spectra
obtained in the present work, it has been possible to identify ISGDR strength
up to higher excitation energies. Second, based on the theoretical work, it is
clear that only the higher-energy component of the observed ISGDR strength needs
to be taken into account for the purposes of extracting the nuclear 
incompressibility. The combination of these two factors leads to the centroid
of the experimental ISGDR strength in
$^{208}$Pb being close to the theoretical predictions of Col\`{o} {\em et al.}\cite{colo} and of Piekarewicz \cite{jorge}.
The first of these employs a value for nuclear incompressibility, K$_{\infty}$
= 215 MeV, and the second uses K$_{\infty}$ = 224 MeV. With both, the GMR
energies are also well reproduced over a range of nuclei. It may be concluded,
then, that a value of K$_{\infty}$ $\approx$ 220 MeV is
consistent with the observed properties of  both the compressional modes in
$^{208}$Pb.

This work has been supported in part by the U. S. National Science Foundation
(grants no. PHY-9901133 and INT-9910015), the University of Notre Dame, and the 
Japan Society for the Promotion of Science (JSPS).

%\vspace{-0.5cm}

\doingARLO[\bibliographystyle{aipproc}]
          {\ifthenelse{\equal{\AIPcitestyleselect}{num}}
             {\bibliographystyle{arlonum}}
             {\bibliographystyle{arlobib}}
          }
\bibliography{berk}

\end{document}